\documentstyle[art12]{article}
\newcommand\sect[1]{\setcounter{equation} 0\section{#1}}

\newcommand{\be}{\begin{equation}}
\newcommand{\ee}{\end{equation}}
\newcommand{\ba}{\begin{eqnarray}}
\newcommand{\ea}{\end{eqnarray}}
\newcommand{\baa}{\begin{eqnarray*}}
\newcommand{\eaa}{\end{eqnarray*}}
\newcommand{\bb}{\begin{thebibliography}}
\newcommand{\eb}{\end{thebibliography}}
\newcommand{\ci}[1]{\cite{#1}}
\newcommand{\bi}[1]{\bibitem{#1}}
\newcommand{\lab}[1]{\label{#1}} 
\newcommand{\re}[1]{(\ref{#1})}
\newcommand{\Tr}{\mbox{Tr\,}}

\newcommand\e{\mbox{e}}
\newcommand\CO{{\cal O}}
\newcommand\CH{{\cal H}}
\newcommand\geff{\bar g}

\newcommand\sq[1]{\sqrt{{#1}^2+c}}
\newcommand\ra[1]{\sqrt{1+\frac{c}{{#1}^2}}}
\newcommand\qqquad{\qquad\quad}
\newcommand\qqqquad{\qquad\qquad}

\newcommand\vev[1]{\langle{#1}\rangle}
\newcommand\VEV[1]{\left\langle{#1}\right\rangle}
\newcommand\bin[2]{\left({#1}\atop{#2}\right)}

\begin{document}
\thispagestyle{empty}
\hfill
\parbox{45mm}{{\sc UPRF--92--340}\par hepth@xxx/9206088 \par May, 1992}

\vspace*{20mm}
\begin{center}
\renewcommand{\thefootnote}{\fnsymbol{footnote}}
{\LARGE Loops in the Curvature Matrix Model}

\vspace{30mm}

\newcommand{\st}{\fnsymbol{footnote}}%

{\large G.~P.~Korchemsky}%
\footnote{On leave from the Laboratory of Theoretical Physics,
          JINR, Dubna, Russia}
\footnote{INFN Fellow}

\bigskip

{\em Dipartimento di Fisica, Universit\`a di Parma and \par
INFN, Gruppo Collegato di Parma, I--43100 Parma, Italy} \par
{\tt e-mail: korchemsky@vaxpr.cineca.it}

\end{center}

\vspace*{20mm}

\begin{abstract}
Macroscopic loop correlators are investigated in the hermitian one matrix
model with the potential perturbed by the higher order curvature term. In
the phase of smooth surfaces the model is equivalent to the minimal
conformal matter coupled to gravity. The properties of the model in the
intermediate phase are similar to that of the discretized bosonic string
with the central charge $C > 1.$ Loop correlators describe the
effect of the splitting of the random surfaces. It is shown, that the
properties of the surfaces are changed in the intermediate phase because
the perturbation modifies the spectrum of the scaling operators.
\end{abstract}

\newpage

\renewcommand{\thefootnote}{\arabic{footnote}}
\setcounter{footnote}{0}

\sect{Introduction}
Computer simulations of the discretized Polyakov's bosonic string
indicate \ci{num}, that the critical behaviour of the model with the
central charge beyond the $C=1$ barrier is governed by the higher order
intrinsic curvature terms in the string action. Therefore trying to write
the matrix model representation of noncritical strings with the central
charge $C > 1$ it seems natural to perturb the potential of the matrix model
by additional terms taking into account the effects of the intrinsic
curvature. One of the possible curvature matrix models was proposed in
\ci{das} and was studied in details \ci{cur,alv}. It was found that the $D=0$
hermitian one matrix model with polynomial potential perturbed by the ``higher
order curvature'' term $(\Tr M^2)^2$ has a phase diagram similar to the
analogous diagram of the discretized Polyakov's bosonic string. It contains the
phase of smooth (Liouville) surfaces, the intermediate phase and the phase
of branched polymers. In the first phase the perturbation becomes
irrelevant and in the continuum limit the model describes nonunitary
minimal $(2,2m-1)$ conformal matter coupled to $2D$ gravity \ci{pol}. The
intermediate phase is the most interesting phase because perturbation
becomes relevant here and the string susceptibility exponent takes positive
values $(\gamma_{str}=1/(m+1)).$ The same property was noticed
\ci{num} for noncritical strings with the central charge $1 < C < 4$
and one hopes that investigation of the curvature matrix model may give
insight into noncritical strings beyond the $C=1$ barrier. However, to make a
correspondence of the curvature matrix model with the continuum theory
one has to study the properties of the intermediate phase in more details.
The calculation of the correlation functions of microscopic and macroscopic
loop operators is the first step in this direction. In the present
paper we perform this calculation at the spherical (genus zero) approximation.

The partition function of the $D=0$ hermitian one matrix model perturbed by
the higher order curvature term is defined as \ci{cur}
\be
\e^{Z(\alpha)}=\int dM\,\exp\left(-\alpha N\Tr V_0(M)
               +\frac14g\alpha^2 (\Tr M^2)^2\right)
\lab{1.1}
\ee
where integration is performed over hermitian $N\times N$ matrices $M.$
Here, $\log\alpha$ is the ``bare'' cosmological constant,
the constant $g$ couples to the perturbation and {\it even\/} potential
$V_0(M)$ is given by
$$
V_0(M)=\int dx\, \rho(x) \log(x-M), \qquad \rho(x)=\rho(-x)
$$
with function $\rho(x)$ being a parameter of the model. The last term
in the exponent of \re{1.1} takes into account effects of the intrinsic
curvature. After expansion of the partition function into sum over random
surfaces generated by the potential $V_0(M),$ this term opens a possibility
for random surfaces to touch each other \ci{das}. In the leading large $N$
(spherical) approximation the factorization property $\vev{(\Tr M^2)^2}=
(\vev{\Tr M^2})^2+\CO(N^0)$ implies that these touchings are measured
effectively with the following constant \ci{cur}
$$
\geff=-g \alpha \VEV{\frac1{N}\Tr M^2}
$$
where $\langle\ldots\rangle$ represents a connected correlator evaluated
with the measure defined in \re{1.1}.

The critical behaviour of the partition function depends on the explicit
form of the function $\rho(x).$ To obtain the phase diagram mentioned
before this function has to satisfy the following equation for all $c$
\be
cf(c)\equiv \int dx\, \rho(x)\left(1-\frac1{\ra{x}}\right)
=\frac2{\alpha_0}\left(1+(m+1)\left(1-\frac{c}{c_0}\right)^m
-(m+2)\left(1-\frac{c}{c_0}\right)^{m+1}\right)
\lab{1.2}
\ee
where $\alpha_0$ and $c_0$ are arbitrary parameters and $m$ is a positive
integer. The calculation of the string susceptibility
$\chi=-N^{-2}d^2Z(\alpha)/d\alpha^2\sim(\alpha-\alpha_{cr})^{-\gamma_{str}}$
shows \ci{cur} that under increasing the touching coupling constant $g$ the
model passes through the following phases.

For $g<g_0=\frac{16}{(\alpha_0c_0)^2}$ the model is in the phase of smooth
(Liouville) surfaces with the string susceptibility exponent
$\gamma_{str}=-1/m,$ $(m=2,3,\ldots).$ The critical values of parameters
are
\be
\alpha_{cr}=\alpha_0, \qqquad c_{cr}=c_0, \qqquad \geff(\alpha_{cr})=0
\lab{1.3}
\ee
where parameter $c<0$ defines the boundary of the cut of one loop
correlator, defined below in \re{1.6} and \re{1.7}.
Near the critical point they scale as
\be
\chi\sim c-c_0\sim (\alpha-\alpha_0)^{1/m}, \qqquad
\geff(\alpha)\sim\alpha-\alpha_0.
\lab{1.4}
\ee
For $g=g_0$ the model turns into the intermediate phase with the critical
exponent $\gamma_{str}=1/(m+1)$. The critical values of parameters are the
same \re{1.3} as in the previous phase, but their scaling is different
\be
\chi\sim \frac1{c-c_0}\sim (\alpha-\alpha_0)^{-1/(m+1)}, \qqquad
\geff(\alpha)\sim(\alpha-\alpha_0)^{m/(m+1)}
\lab{1.5}
\ee
For $g>g_0$ the touching term dominates in \re{1.1}. The random surfaces are
degenerated into branched polymers and the string susceptibility exponent has
a maximum value $\gamma_{str}=1/2.$

Let us calculate the correlation functions of micro- and macroscopic
operators in different phases. They are given by expressions like $\Tr M^{2n}$
for finite and infinite $n,$ respectively. One notes, that the same operators
appear in the asymptotics of one loop correlator $\vev{W(z)}$ for large
$z$%
\footnote{Loop correlator is odd function of $z$ for even potential}
\be
\vev{W(z)}=\VEV{\frac1N\Tr\frac1{z-M}}
=\frac1{z}+\frac1{z^3}\VEV{\frac1N\Tr M^2} + \CO(z^{-5})
=\frac1{z}+\frac1{z^3}\left(-\frac{\geff}{g\alpha}\right) + \CO(z^{-5})
\lab{1.6}
\ee
and we will use $\vev{W(z)}$ as a generating functional of the loop
amplitudes. The general expression for (one-cut) one loop correlator in
the model \re{1.1}
is \ci{cur}
\be
\vev{W(z)} = \frac{\alpha}2\int \frac{dx\,\rho(x)}{z-x}
\left(1-\frac{\sq{z}}{\sq{x}}\right)+\frac{\alpha}2\geff (z-\sq{z})
\lab{1.7}
\ee
Here, the explicit form of $c$ and $\geff$ can be found \ci{cur} by comparing
of this expression with the asymptotic expansion \re{1.6} and their scaling
in different phases near the critical point \re{1.3} is given by \re{1.4}
and \re{1.5}.

\sect{Loop correlators}

In matrix model operators like $\Tr M^{2n}/l$ with $l=na^{2\gamma}$ and
$n\to\infty$ create unmarked holes in the random surface. The boundary loop
has length $l$ in units of the lattice spacing $a$ and the exponent $\gamma$
is fixed by the condition $l=$~fixed as $n\to\infty.$ As one will show below,
$\gamma$ is related to the string susceptibility exponent as $\gamma=
|\gamma_{str}|.$ We use the following definition of the macroscopic loop
operator
$$
w(l)=\lim_{n\to\infty,  a\to 0} \Tr M^{2n}/(-c_0)^n, \qqqquad
l=na^{2|\gamma_{str}|}
$$
To reach the continuum limit one introduces scaling variables following
\ci{dou}. Lattice spacing $a,$ renormalized cosmological constant $t,$
renormalized string coupling constant $\kappa$ and specific heat $u(t)$
are defined as
\be
\frac{\alpha}{\alpha_0}=1+a^2t,  \qquad
Na^{2-\gamma_{str}}=\kappa^{-1}, \qquad
\chi=a^{-2\gamma_{str}}(u(t))^{\mbox{sign}(-\gamma_{str})}
\lab{2.1}
\ee
To find the relation between $u$ and $t$ in different phases (string
equation) at the lowest order in $\kappa$ one substitutes
\re{2.1} into \re{1.4} and \re{1.5}.

To extract  macroscopic loop correlators from \re{1.7} we differentiate the
both
sides of \re{1.7} with respect to $\alpha,$ take into account the dependence
$c=c(\alpha)$ and after some algebra obtain
\be
\frac{d}{d\alpha^{-1}}\left(\alpha^{-1}W(z)\right)
=\frac1{2\sq{z}}\frac{dc}{d\alpha^{-1}}\left((cf(c))'-\frac12\geff\right)
-\frac14\frac{d\geff}{d\alpha^{-1}}\int_{0}^{c}\frac{dx}{\sqrt{z^2+x}}
\lab{2.2}
\ee
where the function $f(c)$ was defined in \re{1.2}. The derivative $dc/d\alpha$
can be easily found from this expression as follows. One expands the both
sides of \re{2.2} into series in $1/z$ for large $z,$ uses asymptotics
\re{1.6} to evaluate the l.h.s.~as $1/z+\CO(z^{-3})$ and then compares
coefficients before $1/z$ to get $dc/d\alpha.$ After substitution of this
derivative, \re{2.2} is replaced by
$$
\frac{d}{d\alpha^{-1}}\left(\alpha^{-1}W(z)\right)=\frac1{\sq{z}}
+\frac14\frac{d\geff}{d\alpha^{-1}}\int_{0}^{c}dx\,x\frac{d}{dx}
\frac1{\sqrt{z^2+x}}
$$
and the comparison of the coefficients before $1/z^{2n+1}$ in the large $z$
limit leads to
\be
\frac{d}{d\alpha^{-1}}\left(\alpha^{-1}\VEV{\frac1N\Tr M^{2n}}\right)
=\bin{2n}{n}\left(-\frac{c}4\right)^n\left(1+\frac{c}4
 \frac{d\geff}{d\alpha^{-1}}\frac{n}{n+1}\right)
\lab{2.3}
\ee
where $\bin{2n}{n}=\frac{(2n)!}{(n!)^2}.$ For $n=0$ this equation becomes
trivial, but for $n=1$ it defines the
derivative $\frac{d\geff}{d\alpha^{-1}}.$ Notice, that for $\geff=0$ it
coincides with an analogous equation in the Kazakov's multicritical model
\ci{kaz}. Now we use equation \re{2.3} to
substitute scaling variables \re{2.1}, to calculate $d\geff/d\alpha$
from \re{1.4} and \re{1.5} and to get in the limit $n\to\infty$ the
differential equation for macroscopic loop correlator $\vev{w(l)}.$
After its integration the macroscopic multiloop
correlators $\vev{w(l_1)\ldots w(l_p)}$ can be derived from
$\vev{w(l)}$ as follows. The identity
$\frac{d}{dx}\frac{\delta}{\delta\rho(x)}Z=-\alpha N\VEV{\Tr\frac1{x-M}},$
following from the definition \re{1.1} of
the model, implies that operator
$-\frac1{\alpha N}\frac{d}{dx}\frac{\delta}{\delta\rho(x)}$
is a generator of multiloop correlators
$\VEV{\prod_{i}\Tr\frac1{x_i-M}}.$
The macroscopic loops are found from these correlators as coefficients in
the expansion into series in $1/x_i.$
Hence, macroscopic loop correlators are
generated from the partition function by the operator
\be
w(l)=-\frac1{\alpha N}\oint_\Gamma\frac{dx}{2\pi i}\,
      \left(-\frac{x^2}{c_0}\right)^n
      \frac{d}{dx}\frac{\delta}{\delta\rho(x)},
\qquad n=la^{-2|\gamma_{str}|} \to \infty
\lab{2.4}
\ee
where integration over contour $\Gamma,$ enclosed singularities of the
integrand, was introduced to extract the proper coefficient. In the
continuum limit the
partition function depends on $\rho(x)$ only through the specific heat $u(t),$
or equivalently through the function $c(t),$ and the derivative in \re{2.4}
acts
effectively as
$\frac{\delta}{\delta\rho(x)}
=\int dt\,\frac{\delta c(t)}{\delta\rho(x)}
 \frac{\delta }{\delta c(t)}.$
The evaluation of the derivative $\frac{\delta c(t)}{\delta\rho(x)}$ is
analogous to that of $\frac{dc}{d\alpha}$ performed before. Namely, one
differentiates the both sides of \re{1.7} with
respect of $\rho(x).$ Then, the expression for
$\frac{\delta W(z)}{\delta\rho(x)}$
depends on two derivatives
$\frac{\delta c}{\delta\rho(x)}$ and $\frac{\delta\geff}{\delta\rho(x)}.$
To find them one compares the first two coefficients in the expansion
of $\frac{\delta W(z)}{\delta\rho(x)}$ for large $z$ with analogous
coefficients fixed by the asymptotics \re{1.6}.
The resulting two equations have a solution
$$
\frac{\delta c}{\delta\rho(x)}=-\frac
{\frac{\delta}{\delta\rho(x)}\left(cf(c)-\frac1{16}g\alpha^2c
 \int_{0}^{c}dy\,y^2f'(y)\right)}
{\left((cf(c))'-\frac12\geff\right)\left(1-\frac1{16}g(\alpha c)^2\right)}
$$
where $c\frac{\delta f(c)}{\delta\rho(x)}=
     -(1+\frac{c}{x^2})^{-1/2}$
follows from the definition \re{1.2}. After expansion of this
expression into series in $1/z^2$ and its substitution into \re{2.4}, the
generator of macroscopic loops is given by
\be
w(l)= \frac{2n}{\alpha N}\bin{2n}{n}\int dt\,\left(\frac{c}{4c_0}\right)^n
\frac1{(cf(c))'-\frac12\geff}
\frac{16-\frac{n-1}{n+1}g(\alpha c)^2}{16-g(\alpha c)^2}
\frac{\delta }{\delta c(t)},
\qquad n=la^{-2|\gamma_{str}|} \to \infty
\lab{2.5}
\ee
Let us apply the expressions \re{2.3} and \re{2.5} to calculate macroscopic
loop
correlators in the different phases.

\subsection{Phase of smooth surfaces}

In this phase one has $\gamma_{str}=-1/m,$ specific heat \re{2.1}
after substitution
into \re{1.4} obeys the string equation
\be
u^m(t)=t,
\lab{2.6}
\ee
function $c(t)$ scales near the critical value \re{1.3} as
$c/c_0=1-a^{2/m}u(t)$ and $d\geff/d\alpha\sim a^0.$
Taking into account the explicit form \re{2.1} of string coupling constant
$\kappa^{-1}=Na^{2+1/m}$ and macroscopic
length $l=na^{2/m},$ we get in the limit $n\to\infty$ from \re{2.3}
the following equation for macroscopic loop $w(l)$
$$
\frac{dw(l)}{dt} = - \frac{\kappa^{-1}}{\sqrt{l}}\e^{-lu(t)}
$$
valid up to unessential factor regular for $g<g_0$
and its solution is
\be
w(l)=\frac{\kappa^{-1}}{\sqrt{l}}\int_{t}^{\infty} dt'\,\e^{-lu(t')}
\lab{2.7}
\ee
The generator \re{2.5} is found analogously using \re{2.1} and \re{1.4} as
\be
w(l)=-\kappa\sqrt{l}\int dt\,\dot{u}(t) \e^{-lu(t)}\frac{\delta }{\delta u(t)}
\lab{2.8}
\ee
To reproduce \re{2.7}, one acts by this operator on the partition function
$Z=-\kappa^{-2}\left(\frac{\partial}{\partial t}\right)^{-2}u(t),$
where
negative powers of derivative denote integration. Applying operator \re{2.8}
to \re{2.7} we find macroscopic two-loop correlator
\be
\vev{w(l_1)w(l_2)}=\sqrt{l_1l_2}\int_{t}^{\infty} dt'\,
                   \dot{u}(t') \e^{-(l_1+l_2)u(t')}
                  =\frac{\sqrt{l_1l_2}}{l_1+l_2}\e^{-(l_1+l_2)u}
\lab{2.9}
\ee
and proceeding further we obtain the multiloop correlators
\be
\vev{w(l_1)\ldots w(l_p)}=\kappa^{p-2}\sqrt{l_1\cdots l_p}
\left(-\frac{\partial}{\partial t}\right)^{p-3}
\left(\dot{u}(t)\e^{-(l_1+\cdots +l_p) u(t)}\right).
\lab{2.10}
\ee
The expressions \re{2.8} and \re{2.10} for multiloop correlators in
the phase of smooth surfaces coincide with analogous expressions in Liouville
theory \ci{moo}.

\subsection{Intermediate phase}

The string susceptibility exponent has positive value in this phase
$\gamma_{str}=1/(m+1).$ The string equation
\be
u^{m+1}(t)=t
\lab{2.11}
\ee
and the scaling of parameters:
$
\frac{c}{c_0}=1-u(t)a^{\frac2{m+1}}
$
and
$
\frac{d\geff}{d\alpha}\sim u^{-1}(t) a^{-\frac2{m+1}}
$
follow from \re{1.5} and \re{2.1}. After substitution of these relations,
of the string coupling constant
$\kappa^{-1}=Na^{2-1/(m+1)}$ and macroscopic
length $l=na^{2/(m+1)}$ into \re{2.3} one gets the following equation
for macroscopic loop
\be
\frac{dw(l)}{dt} = - \frac{\kappa^{-1}}{u\sqrt{l}}\e^{-lu(t)}
\lab{2.12}
\ee
The solution has a form
\be
w(l)=\frac{\kappa^{-1}}{\sqrt{l}}
\int_{t}^{\infty} dt'\,\frac1{u(t')}\e^{-lu(t')}
\lab{2.13}
\ee
and it differs from \re{2.7} only by $u$ in the integrand. Calculation of the
generator \re{2.5} in the intermediate phase requires to be careful because for
$g=g_0$ both numerator and denominator of \re{2.5} vanish as $n\to\infty$
\be
w(l)=-\frac{\kappa}{\sqrt{l}}\int dt\,\dot{u}(t)(1+lu(t))\e^{-lu(t)}
\frac{\delta }{\delta u(t)}
\lab{2.14}
\ee
The macroscopic two-loop correlator is found by applying operator \re{2.14}
to \re{2.13}%
\footnote{The same result was announced in \ci{alv}}
\be
\vev{w(l_1)w(l_2)}=\frac{\sqrt{l_1l_2}}{l_1+l_2}\e^{-(l_1+l_2)u}
+\frac1{u\sqrt{l_1l_2}}\e^{-(l_1+l_2)u}
\lab{2.15}
\ee
This expression differs from
two-loop amplitude \re{2.9} in the Liouville phase
only by last term which is singular for $l_1\to 0$ or $l_2\to 0.$
Using \re{2.12} it can be rewritten as
$$
\kappa^2 u\frac{d}{dt}\vev{w(l_1)}\frac{d}{dt}\vev{w(l_2)}
=\vev{w(l_1)\sigma_0}\frac1{\vev{\sigma_0\sigma_0}}\vev{\sigma_0w(l_2)}
$$
where $\sigma_0=-\frac{\partial}{\partial t}$
is the puncture operator. This expression suggests the following
interpretation of the last term in \re{2.15}: it describes the effect of
the splitting of the random surface into two hemispheres with boundary loops
having lengths $l_1$ and
$l_2.$ We will show in the next section that the form of random surfaces
is changed in the intermediate phase because the touching term changes the
properties of the microscopic state associated to the puncture operator.
Acting by operator \re{2.14} on \re{2.15} one calculates macroscopic
multiloop correlators as
\be
\vev{w(l_1)\ldots w(l_p)}=-\kappa^{p-2}
\left(\frac{\partial}{\partial t}\right)^{p-3}
\left(\dot{u}(t)u^{-2}(t)\pi_{l_1}(u(t))\ldots\pi_{l_p}(u(t))\right)
\lab{2.16}
\ee
where $\pi_{l}(u)=\frac{u^2}{\sqrt{l}}\frac{\partial}{\partial u}
\left(\frac1{u}\e^{-lu}\right).$
The properties of random surfaces described by this expression
are considered in the next section where we analyze the influence of the
perturbation on the spectrum of the model.

\sect{From loops to states}

To make the correspondence with the continuum theory one finds the
spectrum of scaling operators in both phases. In the phase of smooth
(Liouville) surfaces the model describes $(2,2m-1)$ minimal conformal matter
coupled to gravity. As was shown in \ci{moo}, macroscopic loop correlators
\re{2.10} contain all information about the scaling operators in this phase.
It turns out that in the intermediate phase there appears a scaling operator
which does not contribute to the macroscopic loops
$\vev{w(l_1)\ldots w(l_p)}$ and, as a consequence, the spectrum of the model
in this phase differs from the spectrum of Liouville.

The spectrum of scaling operators one finds in a standard way perturbing the
string equations \re{2.6} and \re{2.11} as
\be
t=u^{1/|\gamma_{str}|}(t)+\sum_{k\geq 1}t_k u^k(t).
\lab{3.1}
\ee
Here, the term with the (renormalized) constant $t_k$ appears after one
adds the scaling operator $\sigma_k$ to the potential of the matrix model.
In the phase of Liouville surfaces the explicit form of $\sigma_k$ is well
known \ci{kaz}. Operators $\sigma_k$ are given by a sum of {\it local\/}
operators $\Tr M^{2n}$ for $n\leq k$ and their correlation functions are
described by the KdV flow \ci{KdV}. In the intermediate phase the touching
term, being a nonlocal in matrix $M,$ becomes relevant and one may expect
that some of the scaling operators also may be nonlocal, like
$\Tr M^{2n}\Tr M^{2k}.$ Moreover, we will show in this section
that nonlocal scaling operators do appear in the intermediate phase and
the redundant ``boundary'' operator \ci{bou} is one of them.%
\footnote{Boundary operator measures the boundary length as
          $\vev{\sigma_B w(l)}=l\vev{w(l)}$ and in the matrix model with
          potential $V(M)$ it is given by \ci{bou}:
          $\sigma_B=d\,\Tr V(\lambda M)/d\lambda$ for $\lambda=1.$ In the
          model \re{1.1} the touching term introduces nonlocal
          contribution to $\sigma_B$}

The correlation functions of the scaling operators $\sigma_k$ are found by
differentiation of the partition function over $t_k.$ Using the relation
$\frac{\partial u}{\partial t_j}=-u^j\frac{\partial u}{\partial t}$
following from \re{3.1}, the generator of correlators of the scaling operator
$\sigma_k$ is given by
\be
\sigma_j=\frac{\partial }{\partial t_j}
        =-\int dt\,\dot{u}(t) u^j(t)\frac{\delta}{\delta u(t)}
\lab{3.2}
\ee
and $\sigma_0=-\frac{\partial }{\partial t}
=\frac{\partial }{\partial t_0}$ is the puncture operator.
Scaling operators control the asymptotics of macroscopic loop $w(l)$ for
small $l$ when the hole becomes microscopic \ci{moo,sei}. The explicit
form of the expansion of $w(l)$ one gets from \re{2.8}, \re{2.14} and
\re{3.2} as
$$
w(l)=\kappa\sum_{n=0}^{\infty}\frac{(-1)^n}{n!}\ l^{n+1/2}\sigma_n
$$
in the Liouville phase and
\be
w(l)=-\kappa\sum_{n=0}^{\infty}\frac{(-1)^n}{n!}(n-1)\ l^{n-1/2}\sigma_n
\lab{3.3}
\ee
in the intermediate phase. These expressions have two important distinctions.
Firstly, the contribution of the puncture operator $\sigma_0$ is singular for
$l\to 0$ in the intermediate phase \re{3.3}. It is the puncture operator that
leads to singular small $l$ behaviour of the last term in the
two loop correlator \re{2.15} responsible for the splitting of the random
surface. Secondly, operator $\sigma_1$ disappears from \re{3.3} and, hence,
it does not contribute to the macroscopic loop correlators in the intermediate
phase.

The expectation value of the operator $\sigma_k$ inserted on the random
surface with the boundary length $l$ defines the wave function
$\vev{\sigma_j(u)w(l)}$ associated to this operator \ci{moo,sei}.
Using relations \re{2.7}, \re{2.13} and \re{3.2} we calculate the wave
functions in both phases:
\be
\psi_j(u,l)=\vev{\sigma_j(u)w(l)}
=\kappa^{-1}\sqrt{l}\int_u^\infty dx\,x^j\e^{-lx}
\lab{3.4}
\ee
in the Liouville phase and
\be
\varphi_j(u,l)=\vev{\sigma_j(u)w(l)}
=\frac{\kappa^{-1}}{\sqrt{l}}\int_u^\infty dx\,x^{j-2}(1+lx)\e^{-lx}
\lab{3.5}
\ee
in the intermediate phase. Properties of the wave functions \re{3.4} were
studied in \ci{moo}. It was shown that \re{3.4} are closely related to the
gravity wave functions of Liouville theory in the minisuperspace
approximation. In particular, the wave functions \re{3.4} satisfy the
Wheeler--deWitt constraint \ci{moo}
\be
\CH_L\ \psi_j=-j(j+1)\psi_j+j(j-1)u^2\psi_{j-2}
\lab{3.6}
\ee
where $\CH_L=-\left(l\frac{d}{dl}\right)^2+(ul)^2+\frac14$
is the Liouville hamiltonian in the minisuperspace approximation.
The comparison of the matrix model results with that in
the continuum theory is based on this equation. In the Liouville theory the
wave functions associated to microscopic scaling operators are eigenstates
of the Liouville hamiltonian under the additional condition on the relation
between the specific heat $u$ and Liouville cosmological constant $\mu$
\be
u=\sqrt{\mu}+\ldots
\lab{3.7}
\ee
where dots denote analytical in $\mu$ terms. To reproduce the Liouville
wave functions in the phase of smooth surfaces one has to perform the
following two transformations \ci{moo}. At first, one uses an ambiguity in the
definition of the basis of scaling operators $\sigma_k$ to choose their linear
combinations $\hat{\sigma}_k$ whose wave functions
$\hat{\psi}_j(u,l)=\vev{\hat{\sigma}_j(u)w(l)}$
diagonalyze the hamiltonian
\be
\CH_L\ \hat{\psi}_j(u,l)=-j(j+1)\hat{\psi}_j(u,l)
\lab{3.8}
\ee
Second, one performs analytical transformation of the ``KdV background''
$\{t_j\}$ in order to go to the ``conformal background'' in which the
string equation has a form \re{3.7}. The explicit form of these transformation
was found in \ci{moo}.

It is interesting to note that equation \re{3.6} is invariant under gauge
transformations
\be
\psi_{2j}\to\psi_{2j}
                 +\varepsilon\frac{u^{2j}}{2j+1}\psi_{0},
\qquad
\psi_{2j+1}\to\psi_{2j+1}
                 +\varepsilon\frac{u^{2j}}{2j+3}\psi_{1}
\lab{3.9}
\ee
for an arbitrary $\varepsilon$ and expression \re{3.4} is one of the
elements of the gauge orbit \re{3.9}. At the same time, the solution of
\re{3.8}, having semiclassical asymptotics $\psi_j(u,l)\to 0$ as
$l\to\infty,$ is unique \ci{moo}
\be
\hat{\psi}_j(u,l)\equiv\vev{\hat{\sigma}_j(u)w(l)}
                       =\kappa^{-1}u^{j+1/2}K_{j+1/2}(ul)
\lab{3.10}
\ee
where $K_{j+1/2}(ul)$ is the modified Bessel function and the specific heat
obeys the string equation \re{3.7}.

The wave functions of the scaling operators in the intermediate phase are
given by \re{3.5} for $j\geq 0.$ One would expect that in the continuum limit
the touching term modifies the Liouville hamiltonian and the wave functions
\re{3.5} do not obey simple equations, like \re{3.6} and \re{3.8}.
Nevertheless, one finds after some algebra a remarkable relation for the
wave functions \re{3.5} in the intermediate phase
\be
\CH_L\varphi_j=-j(j-1)\varphi_j+j(j-3)u^2\varphi_{j-2}
\lab{3.11}
\ee
It is invariant under gauge transformations
\be
\varphi_{2j}\to\varphi_{2j}
                 +\varepsilon\frac{u^{2j}}{2j-1}\varphi_{0},
\qquad
\varphi_{2j+1}\to\varphi_{2j+1}
                 +\varepsilon\frac{u^{2j-2}}{2j+3}\varphi_{3}
\lab{3.12}
\ee
for an arbitrary $\varepsilon.$ Equations for the wave functions in
both phases, \re{3.6} and \re{3.11}, look similar. Moreover, it can be
easily checked that their solutions are related as
\be
\varphi_{j}(u,l)=\frac{j}{j-1}\psi_{j-1}(u,l),
\qqqquad j \geq 2
\lab{3.13}
\ee
It is important to recognize, that this relation is not valid for two wave
functions: $\varphi_0$ and $\varphi_1.$ As a consequence, the gauge
transformations of $\psi_j$ and $\varphi_j$ are different and the equation
\re{3.13} is not invariant under \re{3.9} and \re{3.12}. For instance, the
expressions \re{3.4} and \re{3.5} belong to different gauge orbits
and they are related as
$\varphi_{j}=\frac{j}{j-1}(\psi_{j-1}-\frac1ju^{j-1}\psi_0),$ but not
\re{3.13}.

The choice of the basis of scaling operators is ambiguous \ci{moo}. In the
phase of smooth surfaces the ambiguity was fixed by the requirement that the
corresponding wave functions $\hat{\psi}_{j}$ have to satisfy the WdW
equation \re{3.8} in the conformal background \re{3.7}. In the intermediate
phase one has not such a reference point and uses instead the relation
\re{3.13} to fix the basis of scaling operators. Then, equations \re{3.13}
and \re{3.10} lead to
\be
\CH_L\ \hat{\varphi}_j(u,l)=-j(j-1)\hat{\varphi}_j(u,l)
\lab{3.14}
\ee
where the wave function $\hat{\varphi}_j$ is equal to the linear
combination of $\varphi_k$ with $k\leq j.$ However, trying to diagonalyze
equation \re{3.11} we meet the following property.
Equation \re{3.11} is not changed if one allows for index $j$ to take negative
values and defines the corresponding functions $\varphi_j$ using \re{3.5}.
In the limit of small length $l$ these functions are singular formally
$\varphi_{-j}\stackrel{l\to 0}{\sim}l^{-1/2},$ $j>0,$
but one uses gauge ambiguity \re{3.12} to transform them as
$\varphi_{-j}\to\hat{\varphi}_{-j}=\varphi_{-j}-\frac1{j+1}u^{-j}\varphi_0.$
The functions $\hat{\varphi}_{-j}$ are regular for small
$l$ and they cannot be treated as wave functions of local operators
\ci{moo,sei}. Only functions $\varphi_j$ for $j\geq 0,$ $j\neq 1$ having
the asymptotics $\varphi_j\stackrel{l\to 0}{\sim}l^{-|j-1/2|}$ are
microscopic wave functions. Examining \re{3.11} for $j=0,1,\ldots$ we
obtain that the wave functions $\varphi_j$ with $j=0$ and $j\geq 2$ form a
linear space under the action of the Liouville hamiltonian. For $j=1$
equation \re{3.11} is replaced by
$
\CH_L\varphi_1=-2u^2\varphi_{-1}
$
and $\CH_L$ mixes $\varphi_1$ with the functions $\varphi_{2j+1}$ for
negative $j,$ since the function $\hat{\varphi}_{-1}$ obeys
$\CH_L\varphi_{-1}=-2\varphi_{-1}+4u^2\varphi_{-3}$ etc.
Note, that in the Liouville phase the functions $\psi_j$ for positive
and negative $j$ are not mixed under the action of $\CH_L$ in \re{3.6}.
One may try to diagonalyze \re{3.11} choosing
\be
\hat{\sigma}_1=\sigma_1
-\sum_{j=0}^\infty\frac{u^{2j+2}}{2j+1}\sigma_{-2j-1}
\lab{3.15}
\ee
and $\hat{\varphi}_1=\vev{\hat{\sigma}_1w(l)},$
but a careful calculation shows that
$\CH_L\hat{\varphi}_1=-\kappa^{-1}\frac{1+lu}{\sqrt{l}}\e^{-lu}$
indicating that the wave function of the scaling operator $\sigma_1$ is not
a eigenstate of the Liouville hamiltonian.

For $j\geq 2$ the wave functions in different phases are related by
\re{3.13} and using \re{3.10} we choose the solutions of \re{3.14} as
\be
\hat{\varphi}_j(u,l)\equiv\vev{\hat{\sigma}_j(u)w(l)}
                      =\kappa^{-1}u^{j-1/2}K_{j-1/2}(ul),
\qqqquad
j\geq 0,\ j\neq 1
\lab{3.16}
\ee
For $j=0$ equation \re{3.11} implies that $\hat{\varphi}_0$ is the
zero mode of the hamiltonian $\CH_L$
and it coincides with the zero mode \re{3.10} in the Liouville phase.
Although the wave functions of the puncture operator are the same in both
phases $(\hat{\psi}_0(u,l)=u\hat{\varphi}_0(u,l))$
their contributions to the macroscopic loops are different.

Equations \re{3.16} and \re{3.5} define a new basis of scaling operators
$\hat\sigma_j$ connected with the ``old'' KdV basis $\sigma_j$ by the relations
\ba
\hat\sigma_j&=&\pi(-1)^{j-1}2^{j-3/2}\sum_{s=0}^{[j/2]}
\frac{2^{-2s}(j-2s-1)u^{2s}}{s!(j-2s)!\Gamma(s-j+3/2)}\sigma_{j-2s}
\lab{3.17} \\
\sigma_j&=&\frac{j!}{j-1}2^{-j+1/2}\sum_{s=0}^{[j/2]}
\frac{(2j-4s-1)u^{2s}}{s!\Gamma(j-s+1/2)}\hat{\sigma}_{j-2s}
\nonumber
\ea
Two-point correlators of $\hat\sigma_j$ can be obtained from \re{3.17} using
the correlators in the KdV basis
\be
\vev{\sigma_j\sigma_k}=-\frac{\kappa^{-2}}{j+k-1}u^{j+k-1},
\qquad
\vev{\hat{\sigma}_j\hat{\sigma}_k}
=\kappa^{-2}\left(-\frac{\pi}2\right)\frac{u^{2j-1}}{2j-1}\delta_{jk},
\qqquad
(j,k\neq 1)
\lab{3.17a}
\ee
up to terms analytic in $u^2.$ The scaling operators $\sigma_1$ and
$\hat{\sigma}_1$ do not appear in these relations as it should be.
The generator of macroscopic loop \re{3.3} has a simple form in this basis
\be
w(l)=-\kappa\sum_{j=0,\ j\neq 1}^\infty
(-1)^j(2j-1)\ u^{-j+1/2}I_{j-1/2}(ul)\ \hat{\sigma_j}
\lab{3.18}
\ee
where $I_{j-1/2}(ul)$ is the modified Bessel function.
Starting from this relation it is possible to explain the origin of the
last term in \re{2.15}. For two loop correlator equations \re{3.18} and
\re{3.16} lead to
\be
\vev{w(l_1)w(l_2)}=-\sum_{j=0,\ j\neq 1}^\infty
(-1)^j(2j-1)\ I_{j-1/2}(ul_1)K_{j-1/2}(ul_2)
\lab{3.19}
\ee
Here, we have a sum over states corresponding to the scaling operators
(except of $\hat{\sigma}_1$) and the $j-$th term is interpreted as a propagator
of the $j-$th state. Let us compare \re{3.19} with an analogous expansion of
two-loop correlator in the Liouville phase \ci{moo}. After identification of
the states \re{3.13}, macroscopic two loop correlators in both phases differ
only by the contribution of the $(j=0)-$state corresponding to the puncture
operator. In the intermediate phase the contribution of $\hat{\sigma}_0$
contains the function $I_{-1/2}(ul)$ singular as $l\to 0$ whereas in the
Liouville phase it is replaced by the function $I_{1/2}(ul)$
regular for small $l.$ Thus, the wave functions of the puncture operator
coincide in both phases but their propagators are different. We use the
identity $I_{-1/2}(z)-I_{1/2}(z)=\frac{2}{\pi} K_{1/2}(z)=\frac{2}{\pi}
K_{-1/2}(z)$ to rewrite \re{3.19} as
\be
\vev{w(l_1)w(l_2)}=\frac2{\pi}K_{1/2}(ul_1)K_{1/2}(ul_2)
-\sum_{j=1}^\infty
(-1)^j(2j-1)\ I_{j-1/2}(ul_1)K_{j-1/2}(ul_2)
\lab{3.20}
\ee
where the second term is equal to the two loop correlator \re{2.9} in the
Liouville phase \ci{moo}, but the first term with
$K_{1/2}(z)=\sqrt{\pi/(2z)}\e^{-z}$ reproduces the splitting term in
\re{2.15}. One transforms the contribution of the puncture operator
$\hat{\sigma}_0$ to the generator \re{3.18} in an analogous way and
developes the following diagram technics for the calculation of
macroscopic loop correlators \ci{moo}. The state $\hat{\varphi}_j$ associated
to the local operator $\hat{\sigma}_j$ is represented as an insertion of
operator on a hemisphere with boundary length $l$ in fig.~1. The two loop
amplitude \re{3.20} is associated to fig.~2 where the first diagram c
orresponds to the splitting term in \re{2.15}. For multiloop correlators
\re{2.16} we get the representation of fig.~3. The correlation functions
of scaling operators $\hat{\sigma}_j$ $(j\neq 1)$ can be obtained by
``sewing''\ci{moo} fig.~1 into fig.~3. These rules are not applicable
for the operator $\hat{\sigma}_1$ because it does not contribute to the
macroscopic loops \re{3.18}.

We found that the wave functions of local operators \re{3.16} and their
propagators coincide with analogous expressions in the Liouville phase.
However, to identify $\hat{\varphi}_j(u,l)$ as microscopic Liouville wave
functions one has to identify the cosmological constant in the intermediate
phase with the Liouville cosmological constant defined in \re{3.7}. The
relation between $\mu$ and the background $t_0,\ldots,t_{m-1}$ in the
intermediate phase can be found by substituting of \re{3.7} into \re{3.1}.
After this identification we can formulate the properties of the
spectrum of scaling operators in the intermediate phase as follows. In the
continuum limit, the higher order curvature term being introduced into the
Liouville theory changes the propagator of the state associated to the
puncture operator and leads to the appearance of a new scaling operator
$\hat{\sigma}_1$ whose wave function does not obey the WdW constraint
\re{3.14}.

The scaling operator $\hat{\sigma}_1$ has the following interpretation. In the
conformal background \re{3.7}, in the intermediate phase, the cosmological
constant couples to the scaling operator $\hat{\sigma}_{m-1}$ and one uses
\re{3.16} to get
$\vev{\hat{\sigma}_{m-1}w(l)}=-d\vev{w(l)}/d\mu
=\kappa^{-1}u^{m-3/2}K_{m-3/2}(ul).$
Integrating this relation and comparing
$\vev{w(l)}=2\kappa^{-1}l^{-1}u^{m-1/2}K_{m-1/2}(ul)$
with \re{3.18} we obtain the expectation values of the scaling
operators \ci{moo}
\be
\vev{\sigma_{m-1}}=\frac{\pi\kappa^{-2}}{(2m-1)(2m-3)}u^{2m-1}, \qqquad
\vev{\sigma_{m+1}}=-\frac{\pi\kappa^{-2}}{(2m-1)(2m+1)}u^{2m+1}
\lab{3.21}
\ee
and $\vev{\sigma_j}=0$ for $j\neq 1, m-1, m+1.$ The expectation value
$\vev{\hat{\sigma}_1}$ is not fixed in such a way because operator
$\hat{\sigma}_1$ does not contribute to the macroscopic loop \re{3.18}.
One uses instead the relations \re{3.15}, \re{3.17} and \re{3.17a}
to evaluate the correlator $\vev{\hat{\sigma}_1\hat{\sigma}_{m-1}}$ as%
\footnote{The correlators of $\hat{\sigma}_{m-1}$ with operators
          $\sigma_1$ and $\sigma_{-2j-1}$ contain logarithmic
          singularities $\sim\log u$ for odd $m$ which disappear in
          the correlator of $\hat{\sigma}_{m-1}$ with the linear
          combination \re{3.15}.
          That was the reason for the transition from operator $\sigma_1$
          to $\hat{\sigma}_1.$}
$$
\vev{\hat{\sigma}_1\hat{\sigma}_{m-1}}
=\sqrt{\frac\pi2}\frac{\kappa^{-2}}{(m-1)(m-2)}u^{m-1},
\qquad \mbox{for even } m
$$
and $\vev{\hat{\sigma}_1\hat{\sigma}_{m-1}}=0$ for odd $m$ up to
analytical in $u^2$ terms. After substitution of \re{3.7} the
integration of
$-d\vev{\hat{\sigma}_1}/d\mu=\vev{\hat{\sigma}_1\hat{\sigma}_{m-1}}$
leads to
$$
\vev{\hat{\sigma}_1}=-\frac{\sqrt{2\pi}\kappa^{-2}}{(m+1)(m-1)(m-2)}u^{m+1},
\qquad \mbox{for even } m
$$
Thus, $\vev{\hat{\sigma}_1}\neq 0$ for even $m$ and using \re{3.21} one may
form a vanishing linear combination of
$u^m\vev{\hat{\sigma}_1},$ $u^2\vev{\hat{\sigma}_{m-1}}$ and
$\vev{\hat{\sigma}_{m+1}}$ which can be considered as expectation value of
the equations of motions in the continuum theory. As was shown before, in
the intermediate phase operators
$\hat{\sigma}_{m-1}$ and $\hat{\sigma}_{m+1}$ coincide with analogous operators
in the Liouville theory and they can be identified as \ci{moo}
$$
\hat{\sigma}_{m+1} \to -\partial^2\phi+\frac1{4\pi\gamma}\hat{R},
\qquad
u^2\hat{\sigma}_{m-1} \to \frac{\mu}{8\gamma}\e^{\gamma\phi},
\qquad
u^m\hat{\sigma}_1 \to \mu^{m/2}\CO(\phi)
$$
where $\CO(\phi)$ is the higher order curvature term. The wave function
corresponding to the operator $\hat{\sigma}_1$, or equivalently $\CO(\phi),$
does not satisfy WdW equation \re{3.14} and it has logarithmic singularities
$\hat{\varphi}_1\sim -\kappa^{-1}l^{-1/2}\log(lu)$ for small $l$ whereas
the Liouville microscopic wave functions \re{3.4} behave as $l^{-j-1/2}.$
It strongly suggests \ci{moo} that $\CO(\phi)$ is essentially nonlocal
operator in matter and Liouville fields. As a consequence, there is no
microscopic state associated to this operator and the correlation functions,
like $\vev{\hat{\sigma}_1 \ldots},$ have not a diagram representation
similar to figs.~1--3.

\sect{Conclusions}

We investigated micro- and macroscopic loop correlators in the hermitian
one matrix model with the potential perturbed by the higher order curvature
term. Perturbation becomes irrelevant in the phase of smooth surfaces and
loop amplitudes coincide with analogous expressions in the $(2,2m-1)$
minimal conformal field theory coupled to Liouville. In the intermediate
phase with the positive string susceptibility exponent
$\gamma_{str}=1/(m+1)$ loop correlators \re{2.15} and \re{2.16} imply
that the perturbation introduces an instability into the formation of the
random surfaces. The surfaces start to split into a few smaller surfaces
and one suspects that their mean size grows as power of the total area in
contrast with logarithmic behaviour in the phase of smooth surfaces. The
properties of the random surfaces are changed in the intermediate phase
because perturbation modifies the spectrum of the scaling operators. After
identification of the Liouville cosmological constant in \re{3.7} the
spectrums in the both phases are very similar. The only difference is that
new scaling operator $\hat{\sigma}_1$ appears and the propagator of the
state associated to the puncture operator is changed in the intermediate phase.
The scaling operator $\hat{\sigma}_1,$ being nonlocal in Liouville
and matter fields, does not contribute to the macroscopic loops and its
wave function does not satisfy Wheeler-deWitt constraint \re{3.14}.
For even $m$ the expectation value of $\hat{\sigma}_1$ modifies the Liouville
equations of motion by higher order curvature term.

\newcommand\PL{Phys.~Lett.~{}}
\newcommand\NP{Nucl.~Phys.~{}}
\newcommand\PR{Phys.~Rev.~D}
\newcommand\CMP{Comm.~Math.~Phys.~{}}
\newcommand\AP{Ann.~Phys.~(NY)~{}}
\newcommand\RMP{Rev.~Mod.~Phys.~{}}
\newcommand\PTP{Prog.~Theor.~Phys.~{}}
\newcommand\CQG{Class.~Quantum~Grav.~{}}
\newcommand\MPL{Mod.~Phys.~Lett.~{}}
\newcommand\PRept{Phys.~Rept.~{}}
\newcommand\PRL{Phys.~Rev.~Lett.~{}}

\bb{99}
\bi{num} F.David, \NP B257 (1985) 45;
\\       J.Ambj{\o}rn, B.Durhuus and J.Fr{\"o}hlich, \NP B257 (1985) 433;
\\       D.V.Boulatov, V.A.Kazakov, I.K.Kostov and A.A.Migdal,
         \NP B257 (1985) 641; \PL 174B (1986) 87.
\bi{das} S.R.Das, A.Dhar, A.M.Sengupta and S.R.Wadia, \MPL A5 (1990) 1041.
\bi{cur} G.P.Korchemsky, ``{\it Matrix model perturbed by
         higher order curvature terms\/}'', Parma Univ. preprint
         UPRF--92--334 (hepth@xxx/9205014).
\bi{alv} L.Alvarez-Gaum\'{e} and J.L.F.Barb\'{o}n, ``{\it A proposal for
         $D > 1$ strings?\/}'', CERN--TH--6464/92 / FTUAM--9209
         (hepth@xxx/9205010).
\bi{pol} A.M.Polyakov, \MPL A2 (1987) 893;
\\       V.G.Knizhnik, A.M.Polyakov and A.B.Zamolodchikov, \MPL A3 (1988)
         819.
\bi{dou} E.Br\'{e}zin and V.Kazakov, \PL 236B (1990) 144;
\\       D.J.Gross and A.A.Migdal, \PRL 64 (1990) 127;
\\       M.Douglas and S.Shenker, \NP 335B (1990) 635.
\bi{kaz} V.A.Kazakov, \MPL A4 (1989) 2125.
\bi{moo} G.Moore, N.Seiberg and M.Staudacher, \NP B362 (1991) 665.
\bi{sei} N.Seiberg, ``{\it Notes on quantum Liouville theory and quantum
         gravity\/}'', Rutgers preprint RU--90--29.
\bi{KdV} T.Banks, M.R.Douglas, N.Seiberg and S.H.Shenker, \PL 238B (1990)
         279.
\bi{bou} E.Martinec, G.Moore and N.Seiberg, \PL 263B (1991) 190.
\eb
\newpage
\pagestyle{empty}

\begin{center}
{\Large{\bf Figures:}}

\bigskip

\bigskip

\setlength{\unitlength}{1mm}
\begin{picture}(50,30)(-18,-15)
\thicklines
\put(0,0){\oval(10,20)}
\put(0,0){\oval(40,20)[r]}
\put(17,0){$*$}
\put(25,0){$\sigma_j$}
\end{picture}

\parbox{130mm}{{\bf Fig.~1:}
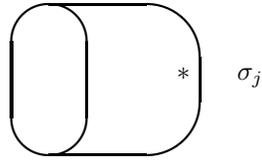
                The wave function associated to the scaling operator
                $\sigma_j.$}

\bigskip

\setlength{\unitlength}{1mm}
\begin{picture}(145,30)(0,-15)
\thicklines
\put(15,0){\oval(10,20)}
\put(60,0){\oval(10,20)[r]}
\thinlines
\put(60,0){\oval(10,20)[l]}
\thicklines
\put(15,0){\oval(40,20)[r]}
\put(60,0){\oval(40,20)[l]}
\put(32,0){$*$}
\put(37,8){$\hat{\sigma}_0$}
\put(41,0){$*$}
\put(1,0){$K_{1/2}$}
\put(66,0){$K_{1/2}\ +\sum_j \  I_{j+1/2}$}
\put(105,0){\oval(10,20)}
\put(135,0){\oval(10,20)[r]}
\thinlines
\put(135,0){\oval(10,20)[l]}
\thicklines
\put(105,10){\line(1,0){30}}
\put(105,-10){\line(1,0){30}}
\put(142,0){$K_{j+1/2}$}
\end{picture}

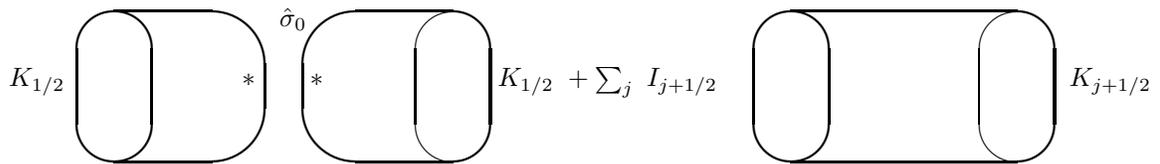
\parbox{130mm}{{\bf Fig.~2:} Macroscopic two-loop amplitude
                $\vev{w(l_1)w(l_2)}$ in the
                intermediate phase. The first diagram describes the splitting
                of the surface.}

\bigskip

\setlength{\unitlength}{1mm}
\begin{picture}(130,50)(-10,-20)
\thicklines
\put(25,-5){\oval(40,20)}
\put(13,20){\oval(6,2.5)}
\put(25,20){\oval(6,2.5)}
\put(37,20){\oval(6,2.5)}
\put(13,20){\oval(6,19)[b]}
\put(13,0){\oval(6,19)[t]}
\put(12,11){$*$}
\put(12,7){$*$}
\put(7,10){$\hat{\sigma}_0$}
\put(22,20){\line(0,-1){20}}
\put(28,20){\line(0,-1){20}}
\put(34,20){\line(0,-1){20}}
\put(40,20){\line(0,-1){20}}
\put(10,24){$K_{1/2}$}
\put(22,24){$I_{j+1/2}$}
\put(34,24){$I_{k+1/2}$}
\put(75,24){$I_{1/2}$}
\put(87,24){$I_{j+1/2}$}
\put(99,24){$I_{k+1/2}$}
\put(-10,-5){$\sum_{j,k}$}
\put(50,-5){$+\ \sum_{j,k}$}
\put(90,-5){\oval(40,20)}
\put(78,20){\oval(6,2.5)}
\put(90,20){\oval(6,2.5)}
\put(102,20){\oval(6,2.5)}
\put(75,20){\line(0,-1){20}}
\put(81,20){\line(0,-1){20}}
\put(87,20){\line(0,-1){20}}
\put(93,20){\line(0,-1){20}}
\put(99,20){\line(0,-1){20}}
\put(105,20){\line(0,-1){20}}
\end{picture}

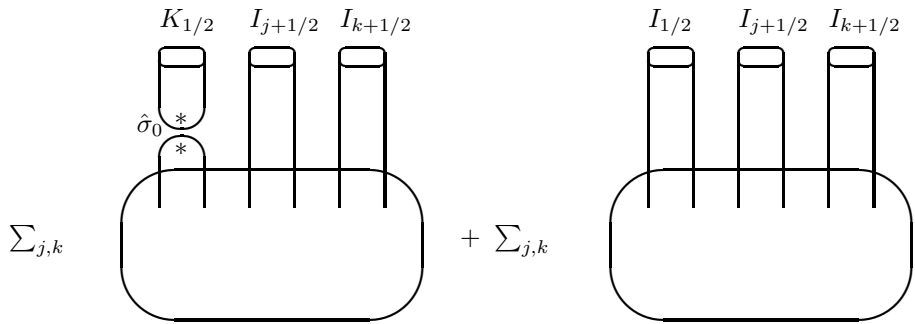
\parbox{130mm}{{\bf Fig.~3:} Some diagrams contributed to the macroscopic
                multiloop amplitude in the intermediate phase.}

\end{center}
\end{document}